\documentclass[12pt]{article}
\usepackage[dvips]{graphicx}
\usepackage{floatflt}
\usepackage{color}
\usepackage{colortbl}

\newcommand{\Pt}{{P_t}}

\newcommand{\dphi}{\Delta\phi}

\newcommand{\gpj}{``$\gamma+jet$''~}

\newcommand{\rrr}{\to} 

\newcommand{\pth}{\hat{p}_{\perp}^{\;min}}

\newcommand{\Ptg}{\Pt^{\gamma}}

\newcommand{\coltab}{0.69}

\newcommand{\lt}{\!<\!}

\newcommand{\hmm}{\hspace*{-1.3mm}}

\newcommand{\Gvc}{\footnotesize{$(GeV/c)$} }

\suppressfloats[!]

\begin{document}

\begin{center}
{\bfseries \gpj EVENT RATE ESTIMATION FOR GLUON DISTRIBUTION DETERMINATION AT THE TEVATRON RUN~II.}

\vskip 5mm

D.V.~Bandurin, N.B.~Skachkov

\vskip 5mm

{\small
{\it
Laboratory of Nuclear Problems, JINR, Dubna, Russia
}
\\
}
\end{center}

\vskip 5mm

\begin{center}
\begin{minipage}{150mm}
\centerline{\bf Abstract}

Since a lot of theoretical predictions on the production of new particles
(Higgs, SUSY) at the Tevatron are based on model estimations of the proton
gluon density behavior at low $x$ and high values of a transfered momentum $Q^2$, 
the study of a possibility of a measurement of the gluon density in this kinematic 
region directly  in Tevatron experiments is obviously of a big interest \cite{D0_Note}. 

\hspace*{5mm} Basing on the selection criteria proposed ealier in 
\cite{D0_Note} and \cite{BKS_P1}, the background events suppression factors and corresponding
signal events selection efficiencies  are determined here. The estimation of the number of 
\gpj events suitable for measurement of gluon distribution in different $x$ and $Q^{\;2}$ 
intervals at Tevatron Run~II is also done.

\hspace*{5mm} It is shown that with integrated luminosity $L_{int}=3~fb^{-1}$
it would be possible to collect about one million of these events. This number
would allow to cover a new kinematical region, $10^{-3}\lt x\lt 1.0$ 
with $1.6\cdot 10^{3}\leq Q^2\leq2\cdot10^{4} ~(GeV/c)^2$, not studied in any previous experiment.
This area includes the values of $Q^2$ that are, on the average, by about one order of magnitude higher 
than those reached at HERA now. 
The rates of $g\,c\to \gamma^{dir} \,+\,jet$ events are also obtained.


\end{minipage}
\end{center}

\section{Selection cuts and background suppression.}                  
To estimate the efficiency of the selection criteria used in \cite{D0_Note}--\cite{QCD_talk2} we
carried out the simulation
\footnote{ PYTHIA~5.7 version with default CTEQ2L parameterization
of structure functions is used here.}
with a mixture of all QCD and SM subprocesses with large cross sections existing in PYTHIA \cite{PYT}
(namely, in notations of PYTHIA, with ISUB=1, 2, 11--20, 28--31, 53, 68). 
The events caused by this set of the subprocesses may give a large background 
to the ``$\gamma^{dir}+jet$'' signal events defined by the subprocesses
(\ref{eq:1a}) and (\ref{eq:1b})
\footnote{A contribution of another possible NLO channel $gg\rrr g\gamma$
(ISUB=115 in PYTHIA) was found to be still negligible even at Tevatron energies.}:
\\[-0.0mm]
``Compton-like'' scattering\\[-10mm]
\begin{eqnarray}
\hspace*{1.04cm} qg\to q+\gamma 
\label{eq:1a}
\end{eqnarray}
\vspace{-6mm}
and  ``annihilation'' process\\[-5mm]
\begin{eqnarray}
\hspace*{1.02cm} q\overline{q}\to g+\gamma  
\label{eq:1b}
\vspace{-3mm}
\end{eqnarray} 
(ISUB=29 and 14) that were also included in the performed simulation.

Three generations  with the above-mentioned set of subprocesses
were done. Each of them was fulfilled with a different value of the PYTHIA parameter
$CKIN(3)\equiv\pth$ that defines the minimal value of parton $\Pt$ appearing in the final state of 
a hard $2\to 2$ parton level fundamental subprocess
in the case of the initial state radiation (ISR) absence. These values were $\pth=40, 70$ and 
$100~GeV/c$.  By 40 million events were generated for each of the $\pth$ value. 
The cross sections of the above-mentioned subprocesses define the rates of the corresponding physical
events and, thus, serve during a simulation as weight factors.

We selected ``$\gamma^{dir}$-candidate +1 jet'' events  containing 
one $\gamma^{dir}$-candidate (denoted in what follows as ${\tilde{\gamma}}$) and one jet 
(found by a subroutine LUCELL implemented in PYTHIA as a jetfinder)
with $\Pt^{jet}> 30~ GeV/c$.
Here and below, as we work at the PYTHIA particle level of simulation, speaking about 
the $\gamma^{dir}$-candidate we actually mean, apart from  $\gamma^{dir}$,  a set of particles
like electrons, bremsstrahlung photons and also photons from neutral meson decays that may be 
registered in one D0 calorimeter
\footnote{The geometry of D0 detector was used in the simulation \cite{D0_det}.}
cell of the $\Delta\eta\times\Delta\phi=0.1\times0.1$ size.

Below we consider a set of 17 cuts that are separated into 2 subsets: a set of the ``photonic'' cuts and 
a set of the ``hadronic'' ones. The first set consists of 5 cuts used to select 
an isolated photon candidate
in some $\Pt^{\tilde{\gamma}}$ interval. The second one, that includes 12 cuts, deals mostly with 
jets and clusters and is used to select events having
one ``isolated jet'' and a limited $\Pt$ activity out of ``${\tilde{\gamma}}+jet$'' system.

All of the cuts are listed in Table ~\ref{tab:sb0}. 
Their influence on the signal-to-background
 ratio $S/B$ is presented in Table 2 (for a case of the most illustrative intermediate interval
of event generation with $\pth=70~GeV/c$) and  Tables 3--5.

Tables 1 and 2 are complementary to each other.
The numbers in the left-hand column (``Cut'') of Table~\ref{tab:sb4}
coincide with the numbers of cuts listed in Table~\ref{tab:sb0}.

The second and third columns of Table 2 contain, respectively, the numbers of 
signal direct photons ($S$) and background $\gamma^{dir}-$candidates ($B$)
left in the sample of events after application of each cut from Table 1. The numbers of
background  events $B$ do not include events with electrons that fake photon signal. 
Their numbers in the samples are presented separately in the last right-hand column ``$e^\pm$''.
The other columns of Table \ref{tab:sb4} include the values of efficiencies
$Eff_{S(B)}$ (with their errors) defined as a ratio
of a number of the signal (background) events that passed under each of the cuts
(1--17) to the number of the preselected events (1st cut of this table).
They are followed by the column containing the values of $S/B$ ratio.
\\[-10mm]
\begin{table}[h]
\small
\caption{\normalsize List of the applied cuts (will be used also in 
Tables \ref{tab:sb4}--\ref{tab:sb3}).}
\begin{tabular}{lc} \hline
\label{tab:sb0}
\hspace*{-2.6mm} {\bf 1}. $a)~\Pt^{\tilde{\gamma}}\geq 40 ~GeV/c, 
~~b)~\Pt^{jet}\geq 30 ~GeV/c,$  
\hspace*{9.99mm} {\bf 9}. $\dphi<17^\circ$; \\
\hspace*{4.1mm} $c)~|\eta^{\tilde{\gamma}}|\leq 2.5, ~~~~~~~~~~d)~\Pt^{hadr}\!<7 ~GeV/c^{\;\ast}$;
\hspace*{6.9mm}{\bf 10}. $\Pt^{miss}/\Pt^{\tilde{\gamma}}\!\leq0.10$; \\
{\bf 2}. $\Pt^{isol}\!\leq 5~ GeV/c, ~\epsilon^{\tilde{\gamma}}<15\%$;
\hspace*{3.12cm} {\bf 11}. $\Pt^{clust}<20 ~GeV/c$; \\
{\bf 3}. $\Pt^{\tilde{\gamma}}\geq\pth$;
\hspace*{5.8cm} {\bf 12}. $\Pt^{clust}<15 ~GeV/c$; \\
{\bf 4}. $\Pt^{isol}_{_ring} \leq 1~ GeV/c^{\;\ast\ast}$;
\hspace*{4.43cm} {\bf 13}. $\Pt^{clust}<10 ~GeV/c$; \\
{\bf 5}. $\Pt^{isol}\!\leq 2~ GeV/c, ~\epsilon^{\tilde{\gamma}}<5\%$;
\hspace*{3.3cm} {\bf 14}. $\Pt^{out}<20 ~GeV/c$; \\
{\bf 6}. $Njet\leq3$;
\hspace*{6.20cm}  {\bf 15}. $\Pt^{out}<15 ~GeV/c$; \\
{\bf 7}. $Njet\leq2$;
\hspace*{6.19cm}  {\bf 16}. $\Pt^{out}<10 ~GeV/c$;\\
{\bf 8}. $Njet=1$;
\hspace*{6.18cm} {\bf 17}. $\epsilon^{jet} \leq 3\%$.\\\hline
\footnotesize{${\;\ast}$ maximal $\Pt$ of a hadron in the ECAL cell containing a 
$\gamma^{dir}$-candidate;}\\
\footnotesize{${\;\ast\ast}$ A scalar sum of $\Pt$ in the ring:
$\Pt^{sum}(R=0.4)-\Pt^{sum}(R=0.2)$.}\\[-3mm]
\end{tabular}
\end{table}
\normalsize

Line number 1 of  Table 1 contains four primary preselection criteria. It includes 
general $\Pt$ cuts (1a), (1b) as well as the cut connected with the D0 electromagnetic
calorimeter (ECAL) geometry (1c) and 
the cut (1d) that excludes $\gamma^{dir}$-candidates accompanied by hadrons entering the same
calorimeter cell as $\tilde{\gamma}$.

Line number 2 of Table 1 fixes the values of $\gamma^{dir}$-candidate isolation parameters
$\Pt^{isol}$ and $\epsilon^{\tilde{\gamma}}$, i.e. 
the value of the scalar sum of $\Pt$ of all particles surrounding
$\gamma^{dir}$-candidate within a cone of $R^{\gamma}_{isol}=((\Delta\eta)^2+(\Delta\phi)^2)^{1/2}=0.7$
and its fraction $\epsilon^{\tilde{\gamma}}=\Pt^{isol}/\Pt^{\tilde{\gamma}}$.

The third cut selects the events with $\gamma^{dir}$-candidate having $\Pt$ higher than
$\pth$ threshold. We impose this cut to select the samples of events with
$\Pt^{\tilde{\gamma}}\geq40, 70$ and $100~GeV/c$ as ISR may smear the sharp kinematical cutoff defined by
$CKIN(3)$ \cite{PYT}. This cut reflects an experimental viewpoint when one is interested in
how many events with $\gamma^{dir}$-candidates are contained in some definite interval of $\Pt^{\tilde{\gamma}}$.

The forth cut restricts a value of $\Pt^{isol}_{_ring}=\Pt^{isol}_{R=0.4}-\Pt^{isol}_{R=0.2}$,
where $\Pt^{isol}_{R}$ is a sum of $\Pt$ of all ECAL cells contained in the cone of the radius 
$R$ around the cell fired by $\gamma^{dir}$-candidate \cite{D0_1}, \cite{D0_2}. 

The fifth cut makes tighter the isolation criteria within $R=0.7$ than those imposed onto 
$\gamma^{dir}$-candidate in the second line of Table 1.
%
%

%
~\\[-24mm]
\begin{table}[htbp]
\begin{center}
\vskip0.0cm
\small
\vspace*{11mm}
\caption{\normalsize Values of significance and  efficiencies for $\pth=70~GeV/c$.}
\vskip0.5mm
\begin{tabular}{||c||c|c|c|c|c|c||}                  \hline \hline
\label{tab:sb4}
Cut& $S$ & $B$ & $Eff_S(\%)$ & $Eff_{B}(\%)$  & $S/B$& $e^\pm$ \\\hline \hline
\rowcolor[gray]{\coltab}%
\rowcolor[gray]{\coltab}%
1 & 39340 &   1247005&  100.00$\pm$  0.00& 100.000$\pm$  0.000&  0.03& 17562\\\hline 
2 & 36611 &     51473 &   93.06$\pm$  0.68 &   4.128$\pm$  0.019 &  0.71 & 4402 \\\hline 
3 & 29903 &     18170 &   76.01$\pm$  0.58 &   1.457$\pm$  0.011 &  1.65  &2038 \\\hline 
4 & 26426 &     11458 &   67.17$\pm$  0.53 &   0.919$\pm$  0.009 &  2.31  &1736 \\\hline 
5 & 23830 &      7504 &   60.57$\pm$  0.50 &   0.602$\pm$  0.007 &  3.18  &1568 \\\hline 
6 & 23788 &     7406 &   60.47$\pm$  0.50 &   0.594$\pm$  0.007  & 3.21  &1554 \\\hline 
7 & 23334 &     6780 &   59.31$\pm$  0.49 &   0.544$\pm$  0.007 &  3.44  &1460 \\\hline 
8 & 19386 &     4136 &   49.28$\pm$  0.43 &   0.332$\pm$  0.005 &  4.69  &1142 \\\hline 
9 & 18290 &      3506 &   46.49$\pm$  0.42 &   0.281$\pm$  0.005  & 5.22  &796 \\\hline 
10 &18022 &      3418 &   45.81$\pm$  0.41 &  0.274$\pm$  0.005  & 5.27  &210 \\\hline 
11 & 15812 &     2600 &   40.19$\pm$  0.38 &   0.208$\pm$  0.004  & 6.08  &176 \\\hline 
12 & 13702 &     1998 &   34.83$\pm$  0.35 &   0.160$\pm$  0.004  & 6.86  &130 \\\hline 
13 & 10724 &     1328 &   27.26$\pm$  0.30 &   0.106$\pm$  0.003  & 8.08  &88 \\\hline 
14 & 10636 &      1302 &   27.04$\pm$  0.30 &   0.104$\pm$  0.003  & 8.17  & 86 \\\hline 
15 & 10240 &      1230 &   26.03$\pm$  0.29 &   0.099$\pm$  0.003  & 8.33  & 84 \\\hline 
\rowcolor[gray]{\coltab}%
16 &  8608 &       984 &   21.88$\pm$  0.26 &   0.079$\pm$  0.003  & 8.75  & 64 \\\hline 
\rowcolor[gray]{\coltab}%
17  & 6266 &       622 &   15.93$\pm$  0.22 &   0.050$\pm$  0.002  &10.07  & 52 \\\hline 
\hline 
\end{tabular}
\end{center}
\vskip-4mm
\noindent
\footnotesize{${(\ast)}$ The background ($B$) does not include
the contribution from the ``$e^\pm$ events'' (i.e. in which $e^\pm$ fake
$\gamma$-candidate) that is shown separately in the right-hand column ``$e^\pm$''.}
\vskip-3mm
\end{table}
\normalsize

The cuts considered up to now, apart from general preselection cut $\Pt^{jet}\geq30~GeV/c$,
used in the first line of Table \ref{tab:sb0},
were connected with photon selection (``photonic'' cuts). 

The cuts 6--9 are connected with the selection of events having only one jet 
and the definition of jet-photon spatial orientation in $\phi$-plane. The 9-th cut selects 
the events with jet and photon transverse momenta being ``back-to-back'' to each other 
in $\phi-$plane within the angle interval of the $\dphi=17^\circ$ size
\footnote{i.e. within the size of three calorimeter cells}.
%

In line 10 we used the cut on $\Pt^{miss}$ to reduce  the background contribution
from the electroweak subprocesses $q\,g \to q' + W^{\pm}$ and $q\bar{~q'} \to g + W^{\pm}$ 
with the subsequent decay $W^{\pm} \to e^{\pm}\nu$ that leads to a substantial $\Pt^{miss}$ value
(see plots in \cite{QCD_talk}). One can see from the last column of Table \ref{tab:sb4} ``$e^\pm$''
that the cut on $\Pt^{miss}$ reduces strongly (in about 4 times) 
the number of events containing $e^\pm$ as direct photon candidate.

Moving further we see from Table \ref{tab:sb4} that the
cuts 11--16 of Table \ref{tab:sb0} reduce the values of transverse momenta of clusters (mini-jets)
$\Pt^{clust}$ and the modulus of a vector sum of all particles that are not included into the
``${\tilde{\gamma}}+ jet$'' system, i.e. $\Pt^{out}$, down to the values less than $10 ~GeV/c$. 
The 17-th cut of Table \ref{tab:sb0} imposes the jet isolation requirement. It leaves only the events
with jets having the sum of $\Pt$ in the ring of the $\Delta R=0.3$ size 
surrounding a jet to be less than $3\%$ of $\Pt^{jet}$.
From comparison of the numbers in the 10-th and 17-th lines we make an important conclusion that all these
new cuts (11--17) may lead to the following about two-fold improvement of $S/B$ ratio
\footnote{One must keep in mind that the starting value of $S/B$ ratio has 
a model dependent nature.}.
This improvement is achieved due to a reduction of $\Pt$ activity out of ``$\tilde{\gamma}+ jet$'' system
\footnote{See \cite{D0_Note} and \cite{BKS_P1} for a detailed definitions of $\Pt^{out}$ and 
jet isolation parameter $\epsilon^{jet}$.}.

It is also important to mention that {\it the total effect of ``hadronic cuts'' 6--17 for the case of $\pth=70~GeV/c$
consists of about twelve-fold decrease of background contribution 
at the cost of less than four-fold loss of signal events (what results in about 3.2 times growth of $S/B$
ratio)}. So, in this sense, we may conclude that from the viewpoint of $S/B$ ratio a study of \gpj events 
may be more preferable as compared with a case of inclusive photon production.

~\\[-14mm]
\normalsize
\begin{table}[h]
\begin{center}
\small
\caption{\normalsize Number of signal and background events remained after cuts.}
\vskip.5mm
\begin{tabular}{||c|c||c|c|c|c|c|c|c||}                  \hline \hline
\label{tab:sb1}
\hmm$\pth$\hmm& &$\gamma$ & $\gamma$ &\multicolumn{4}{c|}{  photons from the mesons}  &
\\\cline{5-8}
\Gvc& Cuts&\hmm direct\hmm &\hmm brem\hmm & $\;\;$ $\pi^0$ $\;\;$ &$\quad$ $\eta$ $\quad$ &
$\omega$ &  $K_S^0$ &\hmm $e^{\pm}$\hmm \\\hline \hline
    &Preselected&\hmm18056&\hmm 14466& 152927& 56379& 17292& 14318&\hmm 2890\hmm  \\\cline{2-9}
 40 &After cuts &\hmm 6238&\hmm 686&     824&  396 &   112& 104&\hmm   24\hmm\\\cline{2-9}
    &+ jet isol. &\hmm 3094&\hmm 264&   338&    150&    40& 44&   14\\\hline  \hline
    &Preselected &\hmm39340&\hmm63982&761926&269666&87932& 63499 &\hmm17562  \hmm\\\cline{2-9}
 70 &After cuts&\hmm  8608 &\hmm  424& 320 &146 & 58  &36 &\hmm 64\hmm \\\cline{2-9}   
    &+ jet isol. &\hmm6266 &\hmm 262 & 206 &90 & 40  & 24 &\hmm 52\hmm \\\hline \hline
    &Preselected&\hmm56764 &\hmm111512 &970710 &346349 &117816 &91416 &\hmm38872\hmm\\\cline{2-9}
100 &After cuts&\hmm 11452 &\hmm 280 & 124 &92 & 24  & 24 &\hmm 136\hmm\\\cline{2-9}
    &+ jet isol. &\hmm 9672 &\hmm 204& 92 & 64 & 24  & 20 &\hmm 120\hmm\\\hline \hline
\end{tabular}
\vskip0.2cm
\caption{\normalsize Efficiency, $S/B$ ratio and significance values in the selected events without jet isolation cut.}
\vskip0.1cm
\begin{tabular}{||c||c|c|c|c|>{\columncolor[gray]{\coltab}}c|c||} \hline \hline
\label{tab:sb2}
$\pth$ \Gvc& $S$ & $B$ & $Eff_S(\%)$  & $Eff_B(\%)$  & $S/B$& $S/\sqrt{B}$
\\\hline \hline
40  & 6238& 2122 & 34.55$\pm$0.51 & 0.831$\pm$0.018 & 2.9 & 135.4 \\\hline
70 & 8608&  984 & 21.88 $\pm$ 0.26 & 0.079 $\pm$ 0.003& 8.8 & 274.4 \\\hline 
100 & 11452& 544 & 20.17 $\pm$ 0.21 & 0.033 $\pm$ 0.001& 21.1 & 491.0 
\\\hline \hline
\end{tabular}
\vskip0.2cm
\caption{\normalsize Efficiency, $S/B$ ratio and significance values in the selected events with jet isolation cut.}
\vskip0.1cm
\begin{tabular}{||c||c|c|c|c|>{\columncolor[gray]{\coltab}}c|c||}  \hline \hline
\label{tab:sb3}
$\pth$ \Gvc& ~~$S$~~ & ~~$B$~~ & $Eff_S(\%)$ & $Eff_B(\%)$  & $S/B$& $S/\sqrt{B}$
 \\\hline \hline
40  & 3094& 836 &17.14$\pm$0.33 & 0.327$\pm$0.011 &  3.7 & 107.0 \\\hline
70 & 6266& 622 & 15.93 $\pm$ 0.22 & 0.050 $\pm$ 0.002& 10.1 & 251.2 \\\hline
100 & 9672& 404 & 17.04 $\pm$ 0.19 & 0.025 $\pm$ 0.001& 23.9 & 481.2 
\\\hline \hline
\end{tabular}
\end{center}
\vskip-3mm
\end{table}
\normalsize

Table \ref{tab:sb1} includes the numbers of signal and background events left in three generated 
event samples after application of cuts 1--16 and 1--17. They are given for all three intervals of 
$\Pt^{\tilde{\gamma}}$.
The summary of Table \ref{tab:sb4} is presented in the middle section ($\pth=70 ~GeV/c$)
of Table \ref{tab:sb1} where the line ``Preselected'' corresponds to the cut 1 of Table \ref{tab:sb0} 
and, respectively, to the line number 1 of  Table  \ref{tab:sb4} presented above.
The line ``After cuts'' corresponds to the line 16 of  Table \ref{tab:sb4} and 
line ``+jet isolation'' corresponds to the line 17 of  Table \ref{tab:sb4}. 

Table \ref{tab:sb1} shows in more detail the origin of $\gamma^{dir}$-candidates.
The numbers in the column ``$\gamma-direct$''  correspond to the numbers  of 
signal events caused by QCD subprocesses (1) and (2)
left in each of $\Pt^{\tilde{\gamma}}$ intervals after application of the cuts defined
in lines 1, 16 and 17 of Table \ref{tab:sb0} (see also column ``$S$'' of Table \ref{tab:sb4}). Analogously,
the numbers in the column ``$\gamma-brem$'' of Table  \ref{tab:sb1} correspond to the numbers
of events with the photons radiated from quarks participating in hard $2\to 2$ parton interactions. 
Columns 5 -- 8 of Table \ref{tab:sb1} illustrate the numbers of the ``$\gamma-mes$''  events with photons
originating from $\pi^0,~\eta,~\omega$ and $K^0_S$ meson decays.
 In a case of $\pth\!=40~GeV/c$ the total numbers of background events,
i.e. a sum over the numbers presented in columns 4 -- 8 of Table \ref{tab:sb1}, 
are shown in lines 1, 16 and 17 of column ``$B$'' of Table \ref{tab:sb4}.

The other lines of Table \ref{tab:sb1} for $\pth\!=40$ and
$~100 ~GeV/c~$ have the meaning analogous to that one described above for $\pth=70 ~GeV/c$.

The last column of Table \ref{tab:sb1} shows the number of preselected events with 
$\gamma^{dir}-$candidates faked by $e^\pm$. One can see that at $\pth=40~GeV/c$ their contribution at the level 
of the last cut ``+jet isol.'' of Table 3 (or cut 17 of Table 1) to the value of the considered total 
background is about $2\%$ and grows up to
$30\%$ at $\pth=100~GeV/c$. Furhter suppression of the events with $\tilde{\gamma}=e^\pm$ depends on 
the value of track finding efficiency.

Here we take the efficiency of each cut of Table 1 to be equal to $100\%$ as we study 
the results of simulation at the particle level. 
For estimation of the size of possible detector effects one may take into account the value of
a track finding efficiency in the central region ($|\eta|\lt 0.9$) of D0 detector 
(\cite{D0_1}, \cite{D0_2}) found in Run~I to be about $83\%$. 

The numbers in Tables \ref{tab:sb2} (without jet isolation cut) and \ref{tab:sb3}
(with jet isolation cut) accumulate in a compact form the final information of 
Tables \ref{tab:sb0} -- \ref{tab:sb1}. 
So, for example,  the values in columns $S$ and $B$ for $\pth=70 ~GeV/c$
include the total numbers of the selected signal and background events taken 
correspondingly at the level of 16-th  (for Table \ref{tab:sb2}) and 17-th (for Table \ref{tab:sb3}) 
cuts from Table \ref{tab:sb4}. 

It is seen from Table \ref{tab:sb2}  that the ratio $S/B$ grows   
from 2.9 to 21.1 while $\Pt^{\tilde{\gamma}}$ increases from
$\Pt^{\tilde{\gamma}}\geq 40 ~GeV/c$ to $\Pt^{\tilde{\gamma}}\geq 100 ~GeV/c$ interval.

The jet isolation requirement (cut 17 from Table \ref{tab:sb0})
noticeably improves the situation at low $\Pt^{\tilde{\gamma}}$ (see Table \ref{tab:sb3}).
After application of this criterion the value of $S/B$ increases from 2.9 
to 3.7 at $\Pt^{\tilde{\gamma}}\geq 40 ~GeV/c$ while at $\Pt^{\tilde{\gamma}}\geq 100 ~GeV/c$ 
the value of $S/B$ changes only from 21.1 to 23.9 due to a tendency of the selected event sample 
to contain more events with an isolated jet as $\Pt^{\tilde{\gamma}}$ increases 
(see \cite{D0_Note} and \cite{BKS_P1}).

Let us underline here that, {\it in contrast to other types of background, the ``$\gamma-brem$'' background
has an irreducible nature}. Thus, the number of ``$\gamma-brem$'' events
should be carefully estimated for each $\Pt^{\tilde{\gamma}}$ interval using the particle level
of simulation in the framework of event generators like PYTHIA.
The contribution of ``$\gamma-brem$'' background must be taken into account in analysis
of the experimental data on prompt photon production  especially 
for high $\Pt$ intervals where, as it is seen
from Table 3, it becomes a dominant one as compared with the contribution from ``$\gamma-mes$''
and ``$e^{\pm}$'' events
\footnote{For a careful theoretical study of `$\gamma-brem$'' background to direct
photon production see \cite{Berg,Aur98} and the corresponding references in them.}.

\normalsize

%
\section{Event rates estimation at the Tevatron.}

One of the promising channels for measurement of the proton gluon density, as was shown in \cite{Au1},
is a high $\Pt$ direct photon production $p\bar{p}(p)\rightarrow \gamma^{dir} + X$.
The region of high $\Pt$, reached by UA1 \cite{UA1}, UA2 \cite{UA2}, CDF \cite{CDF1} and
D0 \cite{D0_1} extends up to $\Pt \approx 60~ GeV/c$ and recently up to $\Pt=105~GeV/c$ \cite{D0_2}. 
These data together with the later ones (see references in \cite{Fer}--\cite{Fr1}) and recent
E706 \cite{E706} and UA6 \cite{UA6} results give an opportunity for tuning the form of gluon 
distribution (see \cite{Au2}, \cite{Vo1}, \cite{Mar}). The rates and estimated cross sections 
of inclusive direct photon production at the LHC are given in \cite{Au1} (see also \cite{AFF}).

Here for the same aim we consider 
the process $p\bar{p}\rightarrow \gamma^{dir}\, +\, 1\,jet\, + \,X$
defined in the leading order by two QCD subprocesses (1) and (2)
(for experimental results see \cite{ISR}, \cite{CDF2} and references in \cite{DMS}). 
The analogous study for LHC energy was done in \cite{BKS_P1},\cite{BKS_GLU} and \cite{DMS}.

The ``$\gamma^{dir}+1 \,jet$'' final state is
more preferable than inclusive photon production $\gamma^{dir} + X$ from
the viewpoint of extraction of information on gluon distribution.
Indeed, in the case of inclusive direct photon production the
cross section is given as an integral over partons distribution
functions $f_a(x_a,Q^2)$ ($a$ = quark or gluon), while in the case of
$p\bar{p}\rightarrow \gamma^{dir}\, +\, 1\,jet\, + \,X$ for $\Pt^{jet}\,
\geq \,30\, GeV/c$ (i.e. in the region where ``$k_t$'' smearing
effects are not important, see \cite{Hu2}) the cross section is
expressed directly in terms of these distributions (see, for example,
\cite{Owe}): \\[-17pt]
\begin{eqnarray}
\frac{d\sigma}{d\eta_1d\eta_2d\Pt^2} = \sum\limits_{a,b}\,x_a\,f_a(x_a,Q^2)\,
x_b\,f_b(x_b,Q^2)\frac{d\sigma}{d\hat{t}}(a\,b\rightarrow 3\,4),
\label{gl:4}
\end{eqnarray}
\vskip-4mm
\noindent
where \\[-9mm]
\begin{eqnarray}
x_{a,b} \,=\,\Pt/\sqrt{s}\cdot \,(exp(\pm \eta_{1})\,+\,exp(\pm \eta_{2})).
\label{eq:x_def}
\end{eqnarray}
We used the following designations above:
$\eta_1=\eta^\gamma$, $\eta_2=\eta^{jet}$; ~$\Pt=\Pt^\gamma$;~ a,b = $q, \bar{q},g$; 
3,4 = $q,\bar{q},g,\gamma$.
Formula (\ref{gl:4}) and the knowledge of the results of independent measurements of
$q, \,\bar{q}$ distributions 
allow the gluon distribution $f_g(x,Q^2)$
to be determined after account of selection efficiencies of $\gamma^{dir}$-candidates 
and the contribution of background, left after the used selection cuts (1--13 of Table \ref{tab:sb0}),
as it was discussed in Section 1 keeping in mind this application.

In the previous works (e.g. see \cite{D0_Note}--\cite{BKS_GLU}) a lot of details connected with the
structure and topology of these events and the objects appearing
in them were discussed. Having all this information we are in position to discuss an application 
of the \gpj event samples selected with the proposed cuts and to estimate the rates of gluon-based  
subprocess (1).

Table~\ref{tab:q/g-1} shows the percentage of ``Compton-like" subprocess (1)
 (amounting to $100\%$ together with (2)) in the samples of generated with PYTHIA events and then 
selected with cuts 1 -- 13 of Table 1 for different $\Ptg$ and $\eta^{jet}$ intervals: 
Central (CC, $|\eta|\lt0.7$), Intercryostat (IC, $0.7\lt|\eta|\lt1.8$) and 
End (EC, $1.8\lt|\eta|\lt2.5$) calorimeter.\\[-7mm]
\begin{table}[h]
\begin{center}
\small
\caption{The percentage of ``Compton-like" process  $q~ g\rrr \gamma +q$.}
\vskip.1cm
\begin{tabular}{||c||c|c|c|c|}                  \hline \hline
\label{tab:q/g-1}
Calorimeter& \multicolumn{4}{c|}{$\Pt^{\gamma}$ interval ($GeV/c$)} \\\cline{2-5}
    part   & 40--50 & 50--70 & 70--90 & 90--140   \\\hline \hline
CC         & 84     &  80   &  74&  68  \\\hline
IC         & 85     &  82   &  76&  70  \\\hline
EC         & 89     &  85   &  82&  73  \\\hline
\end{tabular}
\end{center}
\vskip-5mm
\end{table}
\normalsize

In Table~\ref{tab:q/g-2} we present the $Q^2 (\equiv(\Ptg)^2)$ and $x$ (defined according 
to (\ref{eq:x_def})) distribution of the number of events (for integrated luminosity
$L_{int}=3~fb^{-1}$) that are caused by subprocess (1)
and passed cuts 1 -- 13 of Table 1 with $\Pt^{isol}\lt 4\;GeV/c$ and ${\epsilon}^{\gamma}\lt 7\%$ 
(applied in the line 5)
\footnote{An application of cuts 14--17, as it is seen from Table 2, leads only to $20\%$ improvement
of $S/B$ ratio.}.

\normalsize
The analogous information for events caused by (1) with the charmed quarks 
$g\,c\to \gamma^{dir} \,+\,c~$ is presented in Table~\ref{tab:q/g-3}. 
The simulation of the process $g\,b\to \gamma^{dir}\,+\,b$ $\;$ shows that the rates
for $b$-quark are 8 -- 10 times lower than those for $c$-quark.
~\\[-6mm]
\begin{table}[h]
\begin{center}
\vskip-0.2cm
\caption{Number of~ $g\,q\to \gamma^{dir} \,+\,q$~
events at different $Q^2$ and $x$ intervals for $L_{int}=3~fb^{-1}$.}
\label{tab:q/g-2}
\vskip-0.2cm
\begin{tabular}{|lc|r|r|r|r|r|r|r|}                  \hline
 & $Q^2$ &\multicolumn{6}{c|}{ \hspace{-0.9cm} $x$ values of a parton} &All $x$  \\\cline{3-9}
& $(GeV/c)^2$  &$.001-.005$ & $.005-.01$ & $.01-.05$ &$.05-.1$ & $.1-.5$ &$.5-1.$
&$.001-1.$     \\\hline
&\hmm\hmm 1600-2500\hmm  & 8582  & 56288  &245157  &115870  &203018  &  3647  &632563  \\\hline
&\hmm\hmm 2500-4900\hmm  &  371  & 13514  &119305  & 64412  &119889  &  3196  &320688 \\\hline
&\hmm\hmm 4900-8100\hmm  &    0  &   204  & 17865  & 13514  & 26364  &  1059  & 59007\\\hline
&\hmm\hmm 8100-19600\hmm &    0  &     0  &  3838  &  5623  & 11539  &   548  & 21549\\\hline
\multicolumn{8}{c|}{}&{\bf 1 033 807}\\\cline{9-9}
\end{tabular}
\end{center}
\end{table}
\begin{table}[h]
\begin{center}
\vskip-1.2cm
\caption{Number of~ $g\,c\to \gamma^{dir} \,+\,c$~
events at different $Q^2$ and $x$ intervals for $L_{int}=3~fb^{-1}$.}
\label{tab:q/g-3}
\vskip-0.2cm
\begin{tabular}{|lc|r|r|r|r|r|r|r|}                  \hline
 & $Q^2$ &\multicolumn{6}{c|}{ \hspace{-0.9cm} $x$ values of a parton} &All $x$\\\cline{3-9}
& $(GeV/c)^2$  &$.001-.005$ & $.005-.01$ & $.01-.05$ &$.05-.1$ & $.1-.5$ &$.5-1.$
&$.001-1.$     \\\hline
&\hmm\hmm 1600-2500\hmm  &  264  &  2318  & 21236  & 11758  & 14172  &    58  & 49805 \\\hline
&\hmm\hmm 2500-4900\hmm  &   13  &   332  &  9522  &  6193  &  7785  &    40  & 23885  \\\hline
&\hmm\hmm 4900-8100\hmm  &    0  &     4  &   914  &  1055  &  1648  &    16  &  3637\\\hline
&\hmm\hmm 8100-19600\hmm &    0  &     0  &   142  &   329  &   612  &     8  &  1092 \\\hline
\multicolumn{8}{c|}{}&{\bf 78 419}\\\cline{9-9}
\end{tabular}
\end{center}
\vskip-.5cm
\end{table}

Fig.~1 shows in the widely used $(x,Q^2)$
kinematic plot (see \cite{Sti}) 
what area can be covered by studying the process $q~ g\rrr \gamma +q$.
The number of expected events in this area is given in Table~\ref{tab:q/g-2}.
From this figure and Table~\ref{tab:q/g-2} it becomes clear that at integrated 
\begin{flushright}
\vskip-3.3mm
\parbox[r]{.47\linewidth}
{ \vspace*{0.1cm}
 \hspace{0mm} 
luminosity $L_{int}=3~fb^{-1}$ it would be possible to study the gluon distribution with good statistics 
of \gpj events in the region of small $x$
at $Q^2$ about one order of magnitude higher than now reached at HERA \cite{H1}, \cite{ZEUS}.
It is worth emphasizing that extension of the experimentally reachable
region at the Tevatron to the region of lower $Q^2$ overlapping with the area
covered by HERA would also be of great interest (for instance, it would allow to test 
analytical solutions of DGLAP equations that describe $Q^2$-evolution of structure functions at small 
values of $x$ \cite{Kot}.}
\end{flushright}
\begin{figure}[h]
   \vskip-81.1mm
   \hspace{-1mm} \includegraphics[width=.52\linewidth,height=81mm,angle=0]{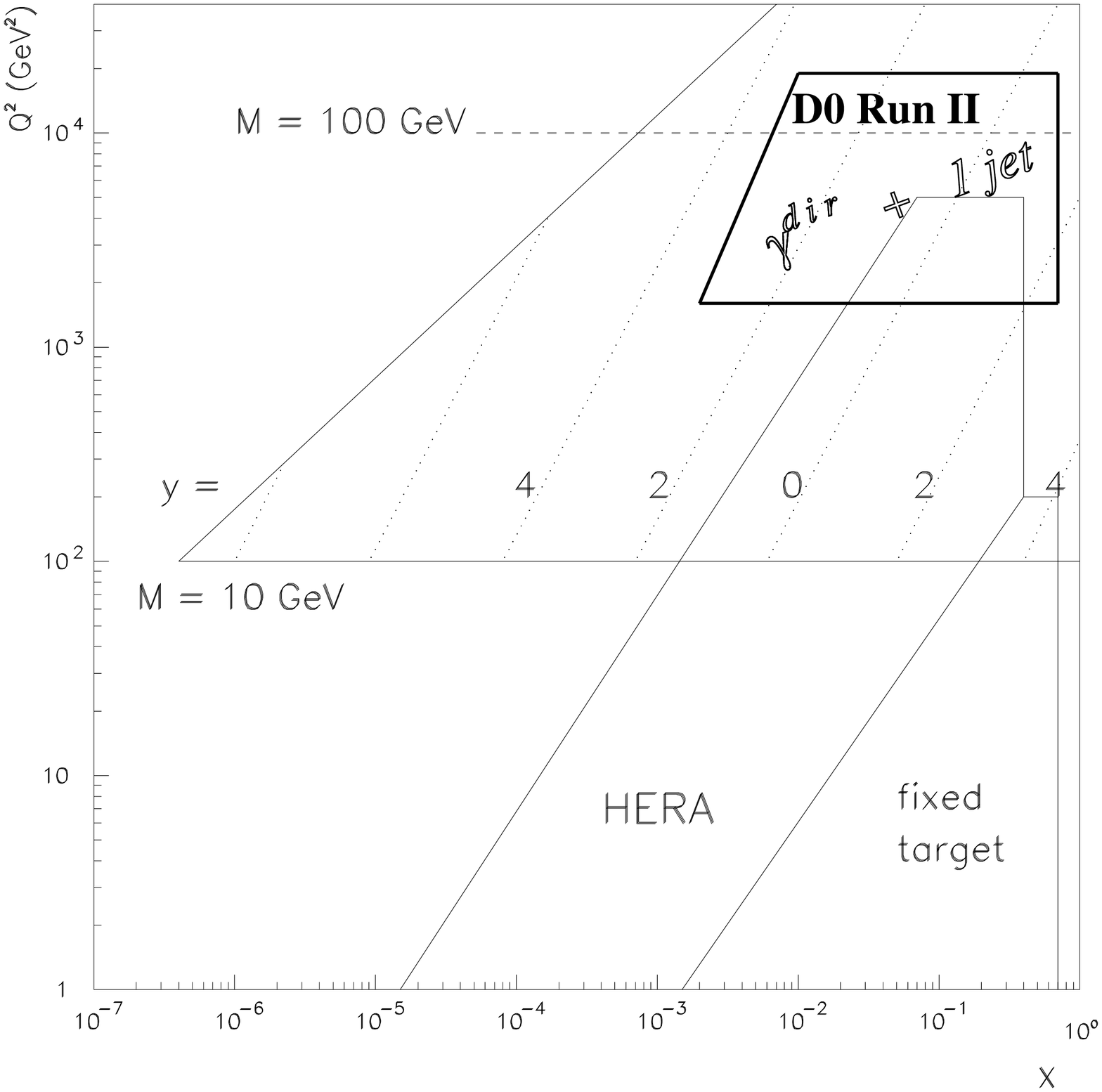}
\label{fig:q/g}
\end{figure}
{\vskip-0.3cm
\hspace*{.5cm} {Figure~1: \normalsize {The  $(x,Q^2)$ kinematic region for
$p\bar{p}\to \gamma^{dir}+jet$ process.}}
}
\newpage

\hspace*{-7mm} \bf{\Large Acknowledgments.}\\[-4pt]  
\normalsize
\rm

We are greatly thankful to D.~Denegri who stimulated us to study the physics of
\gpj processes, permanent support and fruitful suggestions.
It is a pleasure for us to express our recognition for helpful discussions to P.~Aurenche,
M.~Dittmar, M.~Fontannaz, J.Ph.~Guillet, M.L.~Mangano, E.~Pilon,
H.~Rohringer, S.~Tapprogge and especially to H.~Weerts and to J.~Womersley for interest in the work and
encouragements.

\end{document}